\documentclass[12pt]{iopart}

\begin{document}

\title[Generalizations of quantum discord]{Generalizations of quantum discord}

\author{Jianwei Xu}

\address{Key Laboratory for Radiation Physics and Technology, Institute of Nuclear Science and Technology, Sichuan University,
Chengdu 610065, China}
\ead{xxujianwei@yahoo.cn}
\begin{abstract}
The original definition of quantum discord of bipartite states was defined
over projective measurements, in this paper we discuss some
generalizations of it. These generalizations are defined over general measurements, rank-one general measurements or Neumark extension measurements. We investigate the nonnegativity, zero-discord sets of all these quantum discords and some properties about them.

\end{abstract}

\maketitle

\section{Introduction: quantum discord over projective
measurements}

Quantum correlation is one of the most striking features in quantum
many-body systems. Entanglement was widely regarded as nonlocal quantum
correlation and it leads to powerful applications \cite
{Nielson2000,Horodecki2009}. However, entanglement is not the only type of
correlation useful for quantum technology. A different notion of measure,
quantum discord, has also been proposed to characterize quantum correlation
based on quantum measurements \cite{Ollivier2001,Henderson2001}. Quantum
discord captures the nonlocal correlation more general than entanglement, it
can exist in some states even if entanglement does vanish. Moreover, it was
shown that quantum discord might be responsible for the quantum
computational efficiency of some quantum computation tasks \cite
{Datta2008,Lanyon2008,Datta2009}.

Recently, quantum discord has attracted increasing attention. Its evaluation
involves optimization procedure, and analytical expressions are known only
for very few cases \cite{Luo2008,Ali2010,Giorda2010}. A witness of quantum
discord for $2\times n$ states was found \cite{Bylicka2010}, while we have
known that almost all quantum states have nonvanishing quantum discord \cite
{Ferraro2010}. Theoretically, the relations between quantum discord and
other concepts have been discussed, such as Maxwell's demon \cite
{Zurek2003,Brodutch2010}, completely positive maps \cite{Shabani2009}, and
relative entropy \cite{Modi2010}. Also, the characteristics of quantum
discord in some physical models and in information processing have been
studied \cite{Werlang2009,Werlang2010,Wang2010,Soares2010}. An interesting geometric measure of quantum discord was introduced \cite{Dakic2010} and discussed \cite{{Luo2010}}. Very recently, operational interpretations of quantum discord were proposed \cite{Madhok2011,Cavalcanti2011},

The original definition of quantum discord was given over projective measurements. In this paper, we discuss some generalizations
of it. These generalizations will be defined over more extensive measurements than projective measurements. For clarity, we first give some notations which will be used
throughout this paper. Let $H^A,H^B$ be the Hilbert spaces of quantum
systems $A$, $B$, respectively, with $dimH^A=n_A$, $dimH^B=n_B$. $I_A,I_B$
are the identity operators on $H^A$ and $H^B$. The reduced
density matrices of a state $\rho ^{AB}$ on $H^A\otimes H^B$ are $\rho ^A=tr_B\rho^{AB} $, $\rho ^B=tr_A\rho^{AB} .$ For any density operators $\rho ,\sigma $
on a Hilbert space H, the entropy of $\rho $ is $S(\rho )=-tr(\rho \log \rho
)$ $(\log \rho =\log _2\rho )$, the relative entropy is $S(\rho ||\sigma
)=tr(\rho \log \rho )-tr(\rho \log \sigma )$. It is known that $S(\rho
||\sigma )\geq 0$ and $S(\rho ||\sigma )=0$ only if $\rho =\sigma $ (\cite
{Nielson2000}, 11.3.1). The conditional entropy of $\rho ^{AB}$ on $%
H^A\otimes H^B$ (with respect to A) is defined as $S(\rho ^{AB})-S(\rho ^A)$%
. The mutual information of $\rho $ is $S(\rho ^A)+S(\rho ^B)-S(\rho ^{AB})$%
, which is nonnegative and vanishing only when $\rho ^{AB}=\rho ^A\otimes
\rho ^B$ (\cite{Nielson2000}, 11.3.4). A general measurement on $\rho ^{AB}$
is denoted by a set of operators $\Phi =\{\Phi _\alpha \}_\alpha $
satisfying $\sum_\alpha \Phi _\alpha ^{\dagger }\Phi _\alpha =I_{A}\otimes I_{B}$, where $%
\dagger $ means Hermitian adjoint, and $\{\Phi _\alpha \}_\alpha $ operates $%
\rho ^{AB}$ as $\widetilde{\rho ^{AB}}=\sum_\alpha \Phi _\alpha \rho
^{AB}\Phi _\alpha ^{\dagger }.$ When $\Phi _\alpha =A_\alpha \otimes I_B$,
where $A_\alpha $ are operators on $H^A,$ we say $\{A_\alpha \otimes
I_B\}_\alpha $ is a one-sided (with respect to A) general measurement.
Moreover, if $A_\alpha =\Pi _\alpha =|\alpha \rangle \langle \alpha |$ and $%
\{|\alpha \rangle \}_{\alpha =1}^{n_A}$ is an orthonormal basis of $H^A,$ we
call $\{\Pi _\alpha \otimes I_B\}_\alpha $ a one-sided projective
measurement. We sometimes simply write $A_\alpha \otimes I_B$ as $A_\alpha $
by omitting identity operators. We use $\widetilde{\rho ^{AB}}$ to denote
the state whose initial state is $\rho ^{AB}$ and experienced a
measurement, and $\widetilde{\rho ^A}=tr_B\widetilde{\rho ^{AB}},\widetilde{%
\rho ^B}=tr_A\widetilde{\rho ^{AB}}$. When the third system C is concerned,
the notations will be similarly extended to it.

Now recall that the original definition of quantum discord of $\rho ^{AB}$
was defined over projective measurements (on A) as \cite{Ollivier2001}
\begin{equation}
D_A^P(\rho ^{AB})=S(\rho ^A)-S(\rho ^{AB})+\inf_{\{\Pi _\alpha \otimes
I_B\}_\alpha }[\Sigma_{\alpha}p_\alpha S(\widetilde{\rho _\alpha ^B}/p_\alpha )].
\end{equation}
In Eq.(1), inf takes over all projective measurements on A, $\widetilde{%
\rho _\alpha ^B}=tr_A(\Pi _\alpha \rho ^{AB}\Pi _\alpha )$, $p_\alpha =tr_B%
\widetilde{\rho _\alpha ^B}$.

Using the joint entropy theorem (\cite{Nielson2000}, 11.3.2),  Eq.(1) can also be written as \cite{Ollivier2001}
\begin{equation}
D_A^P(\rho ^{AB})=S(\rho ^A)-S(\rho ^{AB})+\inf_{\{\Pi _\alpha \otimes
I_B\}_\alpha }[S(\widetilde{\rho ^{AB}})-S(\widetilde{\rho ^A})].
\end{equation}

A state $\rho ^{AB}$ satisfying $D_A(\rho ^{AB})=0$ is called classical state, it
can be proved \cite{Ollivier2001}
\begin{equation}
D^{P}_A(\rho ^{AB})=0\Longleftrightarrow \rho ^{AB}=\sum_{\alpha
=1}^{n_A}p_\alpha |\alpha \rangle \langle \alpha |\otimes \rho _\alpha ^B,
\end{equation}
where, $\left\{ |\alpha \rangle \right\} _{\alpha =1}^{n_A}$ is an arbitrary
orthonormal set of $H^A$,  $p_\alpha \geq 0$ , $\sum_\alpha p_\alpha =1$. $\rho _\alpha ^B$
are density operators on $H^B$.

Eq.(1) or Eq.(2) have intuitive physical meanings, namely, $D^{P}_A(\rho ^{AB})$ is the minimal loss of conditional entropy or mutual information over all projective measurements. To generalize the definition of quantum discord to other measurements, a direct idea is, we define the quantum discord as Eq.(1) or Eq.(2) but let inf take other measurements. Doing this, we must guarantee the nonnegativity of the definitions like Eq.(1) or Eq.(2) since the positive quantum discord was regarded as a measure of quantum correlation.

The remainder of this paper is arranged as follows. In Sec.II, we consider the
generalization of Eq.(2) to general measurements. In Sec.III, we consider the generalization of Eq.(1) to general measurements and the quantum discord defined over Neumark extension measurements.
 In Sec.IV, we discuss some relations and properties about these quantum discords.
 Finally, Sec.V is devoted to a brief summary.

\section{Generalization of Eq.(2) to general measurements}

To generalize Eq.(2) to general measurements, we  first prove the theorem below.

\textit{Theorem.} For any state $\rho ^{AB}$ and any general measurement $\{A_\alpha \otimes I_B\}_\alpha $ performing on A, it holds that
\begin{equation}
S(\rho ^A)-S(\rho ^{AB})+[S(\widetilde{\rho ^{AB}})-S(\widetilde{\rho ^A})]\geq 0.
\end{equation}

\textit{Proof.} Suppose
\begin{equation}
\rho ^{AB}=\sum_ic_i\rho _i^A\otimes \rho _i^B,
\end{equation}
where $\rho _i^A$, $\rho _i^B$ are Hermitian operators on $H^A$ and $H^B$, $c_i$ are real numbers (this is a very useful representation for bipartite states, see, e.g., \cite{Schlienz1995}). Performing a general measurement $\{A_\alpha \otimes I_B\}_\alpha $ on $\rho ^{AB}$, we have
\numparts
\begin{eqnarray}
&&\rho ^A=tr_B\rho ^{AB}=\sum_ic_i\rho _i^A(tr_B\rho _i^B),  \\
&&\rho ^B=tr_A\rho ^{AB}=\sum_ic_i(tr_A\rho _i^A)\rho _i^B,  \\
&&\widetilde{\rho ^{AB}}=\sum_\alpha A_\alpha \rho ^{AB}A_\alpha ^{\dagger }=\sum_{\alpha i}c_iA_\alpha \rho _i^AA_\alpha ^{\dagger }\otimes \rho _i^B, \\
&&\widetilde{\rho _\alpha ^A}=tr_B(A_\alpha \rho ^{AB}A_\alpha ^{\dagger })=A_\alpha \rho ^AA_\alpha ^{\dagger },\\
&&\widetilde{\rho ^A}=\sum_\alpha \widetilde{\rho _\alpha ^A}=\sum_\alpha A_\alpha \rho ^AA_\alpha ^{\dagger }, \\
&&\widetilde{\rho _\alpha ^B}=tr_A(A_\alpha \rho ^{AB}A_\alpha ^{\dagger })=\sum_ic_i[tr_A(A_\alpha \rho _i^AA_\alpha ^{\dagger })]\rho _i^B, \\
&&\widetilde{\rho ^B}=\sum_\alpha \widetilde{\rho _\alpha ^B}=\rho ^B, \\
&&tr_A\widetilde{\rho _\alpha ^A}=tr_A(A_\alpha \rho ^AA_\alpha ^{\dagger })=tr_B\widetilde{\rho _\alpha ^B}, \\
&&\widetilde{\rho ^A\otimes \frac{I_B}{n_B}}=\sum_\alpha A_\alpha \rho ^AA_\alpha ^{\dagger }\otimes \frac{I_B}{n_B}=\sum_{\alpha}\widetilde{\rho _\alpha ^A}\otimes \frac{I_B}{n_B}.
\end{eqnarray}
\endnumparts
In Eq.(6g), we have used $\sum_\alpha tr_A(A_\alpha ^{\dagger }A_\alpha )=I_A$.
Notice that $\widetilde{\rho ^{AB}}$, $\widetilde{\rho ^A}$, $\widetilde{%
\rho ^B}$ are all density operators, but $\widetilde{\rho _\alpha ^A}$,
$\widetilde{\rho _\alpha ^B}$, $\rho _i^A$, $\rho _i^B$ are not necessarily
so.

For any density operators $\rho ^{AB}$ and $\sigma ^{AB}$, and any general
measurement $\Phi =\{\Phi _\mu \}_\mu $, the monotonicity of relative
entropy reads \cite{Lindblad1975}
\begin{equation}
S(\Phi \rho ^{AB}||\Phi \sigma ^{AB})\leq S(\rho ^{AB}||\sigma ^{AB}).
\end{equation}
Also, conditional entropy and relative entropy have the relation (\cite
{Nielson2000}, 11.4.1)
\begin{equation}
S(\rho ^{AB}||\rho ^A\otimes \frac{I_B}{n_B})=S(\rho ^A)-S(\rho ^{AB})+\log
n_B.
\end{equation}
Now, letting $\sigma ^{AB}=\rho ^A\otimes \frac{I_B}{n_B}$ and $\Phi =\{A_\alpha \otimes I_B\}_\alpha$ in Eq.(7),
combining Eq.(6i) and Eq.(8), we can surely get Eq.(4), then end this proof.

We denote the set of all general measurements on A by G,
\begin{eqnarray}
G=\{\{A_{\alpha}\}_{\alpha}:\sum_\alpha A_\alpha ^{\dagger }A_\alpha =I_A\},
\end{eqnarray}
and denote the set of all rank-1 general measurements on A by R,
\begin{eqnarray}
\fl
R=\{\{\frac{|\gamma \rangle \langle \gamma |}{\sqrt{p_\gamma }}\}_{\gamma
=1}^n:|\gamma \rangle \in H^A,
\sum_\gamma |\gamma \rangle \langle \gamma|=I_A,
p_\gamma =\langle \gamma |\gamma \rangle;all \ n,n\geq n_A\}.
\end{eqnarray}

Now from Eq.(2) and Eq.(4), we define
\begin{eqnarray}
D_A^S(\rho ^{AB})=S(\rho ^A)-S(\rho ^{AB})+\inf_{S\otimes I_B}[S(\widetilde{\rho ^{AB}})-S(\widetilde{\rho ^A})],
\end{eqnarray}
where $S\subset G$. Under this definition, Proposition 1, Proposition 2 and Proposition 3 below are easy to get.

\textit{Proposition 1.} For any state $\rho ^{AB}$, $D_A^S(\rho ^{AB})$ defined as in Eq.(11), then
\begin{eqnarray}
D_A^S(\rho ^{AB})\geq 0.
\end{eqnarray}

\textit{Proposition 2.} For any state $\rho ^{AB}$, $D_A^S(\rho ^{AB})$ defined as in Eq.(11), then
\begin{eqnarray}
\{I_A\}\in S \Rightarrow D_A^S(\rho ^{AB})= 0.
\end{eqnarray}

Eq.(13) is true, because  $\{I_A\}$ results in the equality in Eq.(4) for any state $\rho ^{AB}$. Since $\{I_A\}\in G$, thus

\textit{Proposition 3.} For any state $\rho ^{AB}$, $D_A^G(\rho ^{AB})$ defined as in Eq.(11), then
\begin{eqnarray}
 D_A^G(\rho ^{AB})= 0.
\end{eqnarray}

So, $D_A^S(\rho ^{AB})$ is not trivial only if $\{I_A\}\notin S$, such as S takes the set P (all projective measurements), or the set R (all rank-1 general measurements).

The intuitive meaning of Eq.(11) is that $D^{S}_A(\rho ^{AB})$ is the minimal
loss of conditional entropy or mutual information (since $\rho ^B=\widetilde{\rho ^B}$, see Eq.(6g)) over a set of some general
measurements on A.

The optimization of Eq.(11) is not an easy thing in general ($I_A\notin S$), but we would like to give an upper bound of it
(although, any general measurement in the set S will yield a
corresponding upper bound). Actually, the mutual information of $\rho ^{AB}$
is an upper bound of $D^{S}_A(\rho ^{AB})$ for any S and any state $\rho ^{AB}$.

\textit{Proposition 4.} For any state $\rho ^{AB}$ and any set $S\in G$, $D_A^S(\rho ^{AB})$ defined as in Eq.(4), it holds that
\begin{eqnarray}
 D_A^S(\rho ^{AB})\leq S(\rho ^A)+S(\rho ^B)-S(\rho ^{AB}).
\end{eqnarray}

 To make clear this assertion, note
that mutual information is nonnegative and $\rho ^B=\widetilde{\rho ^B}$ (see Eq.(6g)),
then $S(\widetilde{\rho ^{AB}})-S(\widetilde{\rho ^A})\leq S(\rho ^B)$.
Moreover, it is known that there exists a set of unitary matrices $U_j$ on $%
H^A$ and probabilities $p_j$ such that $\sum_jp_jU_j\otimes I_B\rho ^{AB}$ $%
U_j^{\dagger }\otimes I_B=\frac{I_A}{n_A}\otimes \rho ^B$ (\cite{Nielson2000}, 11.3.4). The measurement $\{\sqrt{p_j}U_j\otimes I_B\}_j$ exactly
achieves $S(\widetilde{\rho ^{AB}})-S(\widetilde{\rho ^A})=S(\rho ^B).$ This implies the equality in Eq.(15) can be achieved for some set S.

\section{Generalization of Eq.(1) to general measurements}
	
We now consider the generalization of Eq.(1) to  general measurements as
\begin{eqnarray}
\overline{D}_A^S(\rho ^{AB})=S(\rho ^A)-S(\rho ^{AB})+\inf_{S\otimes I_B}[\sum_\alpha p_\alpha S(\widetilde{\rho _\alpha ^B}/p_\alpha )],
\end{eqnarray}	
where $S\subset G$, $p_\alpha =tr_B\widetilde{\rho _\alpha ^B}$, $\widetilde{\rho _\alpha ^B}$ specified in Eq.(6f). We need to prove $\overline{D}_A^S(\rho ^{AB})\geq 0$. To do this, we first point out that
\begin{eqnarray}
[\sum_\alpha p_\alpha S(\widetilde{\rho _\alpha ^B}/p_\alpha )]_{\{A_\alpha \}_\alpha }\geq [\sum_\alpha p_\alpha S(\widetilde{\rho _\alpha ^B}/p_\alpha )]_{R\{\{A_\alpha \}_\alpha \}}.
\end{eqnarray}
 In Eq.(17), $[\sum_\alpha p_\alpha S(\widetilde{\rho _\alpha ^B}/p_\alpha )]_{\{A_\alpha \}_\alpha }$ means $[\sum_\alpha p_\alpha S(\widetilde{\rho _\alpha ^B}/p_\alpha )]$ was defined under the general measurement $\{A_\alpha \}_\alpha $, and $R\{\{A_\alpha \}_\alpha \}$ is the rank-1 decomposition of $\{A_\alpha \}_\alpha $,
\begin{eqnarray}
R\{\{A_\alpha \}_\alpha \}=\{\{\frac{|\alpha _j\rangle \langle \alpha _j|}{\sqrt{\langle \alpha _j|\alpha _j\rangle }}\}_{\alpha ,\alpha _j}:A_\alpha ^{\dagger }A_\alpha =\sum_{\alpha _j}|\alpha _j\rangle \langle \alpha _j| \}.
\end{eqnarray}
 In Eq.(18), $A_\alpha ^{\dagger }A_\alpha =\sum_{\alpha _j}|\alpha _j\rangle \langle \alpha _j|$ is the eigendecomposition of the positive operator $A_\alpha ^{\dagger }A_\alpha$.
Eq.(17) can be obtained \cite{Devetak2004,Koashi2004} by using the concavity of entropy (\cite{Nielson2000}, 11.3.5)
\begin{eqnarray}
S(\sum_ip_i\rho _i)\geq \sum_ip_iS(\rho _i),
\end{eqnarray}
where $p_i\geq 0$,  $\sum_ip_i=1$, $\rho _i$ are density operators.

Here we would like to consider another generalization of quantum discord, which defined over Neumark extension measurements.

Neumark extension \cite{Neumak1940,Peres1995} says any general measurement $\{\frac{|\gamma \rangle \langle \gamma |}{\sqrt{p_\gamma }}\}_{\gamma =1}^n$ of R can be extended to a projective
measurement $\{|\overline{\gamma }\rangle \langle \overline{\gamma }%
|\}_{\gamma =1}^n$ on $H_n^A$, here $H_n^A$ is a direct-sum extended Hilbert
space of $H^A$ with $dimH_n^A=n\geq n_{A}$, and $|\gamma \rangle $ are just the restrictions of $%
|\overline{\gamma }\rangle $ onto $H^A$. Evidently, the Neumark extension $%
\{|\overline{\gamma }\rangle \langle \overline{\gamma }|\}_{\gamma =1}^n$
for $\{\frac{|\gamma \rangle \langle \gamma |}{\sqrt{p_\gamma }}\}_{\gamma
=1}^n$ is not necessarily unique, but given any orthonormal basis $\{|%
\overline{\gamma }\rangle \}_{\gamma =1}^n$ of $H_n^A$, there is only one $\{%
\frac{|\gamma \rangle \langle \gamma |}{\sqrt{p_\gamma }}\}_{\gamma =1}^n$
that $\{|\overline{\gamma }\rangle \langle \overline{\gamma }|\}_{\gamma
=1}^n$ is its Neumark extension. (Some recent discussions about Neumark extension see \cite{Luis2002,Andersson2008,Samsonov2009}.)
Let
\begin{eqnarray}
\fl
N=\{\{|\overline{\gamma }\rangle \langle \overline{\gamma }|\}_{\gamma
=1}^n:\{|\overline{\gamma }\rangle \}_{\gamma =1}^n \ is \ any \ orthonormal \ basis \ of \ H_n^A;all \ n, n\geq n_A\}.
\end{eqnarray}

We now consider the quantum discord $D_A^{N}(\rho ^{AB})$ over Neumark
extension measurements as
\begin{equation}
\overline{\overline{D}}_A^N(\rho ^{AB})=S(\rho ^A)-S(\rho ^{AB})+\inf_{N\otimes I_B}[S(\widetilde{
\rho ^{AB}})-S(\widetilde{\rho ^A})].
\end{equation}
 Just as the equivalence of Eq.(1) and Eq.(2), Eq.(21) can also be written as the form of Eq.(1).

 Suppose a general measurement $\{\frac{|\gamma \rangle \langle \gamma |}{\sqrt{p_\gamma }}\}_{\gamma =1}^n$ of R, $\{|\overline{\gamma }\rangle \langle \overline{\gamma }|\}_{\gamma=1}^n$ is its Neumark extension, note that  $\{\frac{|\gamma \rangle \langle \gamma |}{\sqrt{p_\gamma }}\}_{\gamma =1}^n$ is performed on $H^A$, while $\{|\overline{\gamma }\rangle \langle \overline{\gamma }|\}_{\gamma=1}^n$ is performed on $H_n^A$.

 Neumark extension measurements are very similar with the projective measurements only performing on a larger space. Since Eq.(4) is hold for a projective measurement, with the similar expressions of Eqs.(6a-6i) for Neumark extension measurements, we steadily get
\begin{eqnarray}
S(\rho ^A)-S(\rho ^{AB})+[\sum_\alpha p_\alpha S(\widetilde{\rho _\alpha ^B}/p_\alpha )]_{\{|\overline{\gamma }\rangle \langle \overline{\gamma }|\}_{\gamma =1}^n}\geq 0.
\end{eqnarray}
	
From Eq.(6f), it is easy to find
\begin{eqnarray}
[\sum_\alpha p_\alpha S(\widetilde{\rho _\alpha ^B}/p_\alpha )]_{\{\frac{|\gamma \rangle \langle \gamma |}{\sqrt{p_\gamma }}\}_{\gamma =1}^n}=[\sum_\alpha p_\alpha S(\widetilde{\rho _\alpha ^B}/p_\alpha )]_{\{|\overline{\gamma }\rangle \langle \overline{\gamma }|\}_{\gamma =1}^n},
\end{eqnarray}
where $\{|\overline{\gamma }\rangle \langle \overline{\gamma }|\}_{\gamma =1}^n$ is the Neumark extension of rank-1 general measurement $\{\frac{|\gamma \rangle \langle \gamma |}{\sqrt{p_\gamma }}\}_{\gamma =1}^n.$

Combining Eq.(17), Eq.(23), Eq.(22), Eq.(16), we obtain proposition 5 and proposition 6 below.

\textit{Proposition 5.} For any state $\rho ^{AB}$, $\overline{D}_A^S(\rho ^{AB})$ defined as in Eq.(16), then
\begin{eqnarray}	
\overline{D}_A^S(\rho ^{AB})\geq 0,	
\end{eqnarray}

\textit{Proposition 6.} For any state $\rho ^{AB}$, $\overline{D}_A^S(\rho ^{AB})$ defined as in Eq.(16), then
\begin{eqnarray}
\overline{D}_A^G(\rho ^{AB})=\overline{D}_A^R(\rho ^{AB})=\overline{\overline{D}}_A^N(\rho ^{AB}).
\end{eqnarray}

Eq.(23) and Eq.(25) will be used frequently in next section.

\section{Some properties about different quantum discords}
We prove that

\textit{Proposition 7.} The quantum discords $D_A^P(\rho ^{AB})$, $D_A^R(\rho ^{AB})$, $\overline{D}_A^R(\rho ^{AB})$ of a bipartite state $\rho ^{AB}$ defined in Eq.(1), Eq.(11) and Eq.(16), hold that
\begin{equation}
D_A^P(\rho ^{AB})\geq D_A^R(\rho ^{AB})\geq \overline{D}_A^R(\rho ^{AB}).
\end{equation}

\textit{Proof}. We only need to prove $D_A^R(\rho ^{AB})\geq \overline{\overline{D}}_A^N(\rho ^{AB})$. First note that for any $\{|\overline{\gamma }\rangle \langle \overline{\gamma }|\}_{\gamma =1}^n\in N$, when restrict it onto $H^A$, we obtain $\{\frac{|\gamma \rangle \langle \gamma |}{\sqrt{p_\gamma }}\}_{\gamma =1}^n\in R$, where $|\gamma \rangle $ is the projection of $|\overline{\gamma }\rangle $ onto $H^A.$ According to the definitions of $D_A^R(\rho ^{AB})$ and $\overline{\overline{D}}_A^N(\rho ^{AB})$, we have
\begin{eqnarray}
\fl
D_A^R(\rho ^{AB})=S(\rho ^A)-S(\rho ^{AB})+\inf_{R\otimes I_B}[S(\sum_{\gamma =1}^n\frac 1{p_\gamma }|\gamma \rangle \langle \gamma |\rho ^{AB}|\gamma \rangle \langle \gamma |)-S(\sum_{\gamma =1}^n\frac 1{p_\gamma }|\gamma \rangle \langle \gamma |\rho ^A|\gamma \rangle \langle \gamma |)], \nonumber\\
\fl
\overline{\overline{D}}_A^N(\rho ^{AB})=S(\rho ^A)-S(\rho ^{AB})+\inf_{N\otimes I_B}\sum_{\gamma =1}^n\langle \overline{\gamma }|\rho ^A|\overline{\gamma }\rangle S(\frac{ \langle \overline{\gamma }|\rho ^{AB}|\overline{\gamma }\rangle }{\langle \overline{\gamma }|\rho ^A|\overline{\gamma }\rangle }). \nonumber
\end{eqnarray}
Then from the concavity of conditional
entropy (\cite{Nielson2000}, 11.4.1), we have
\begin{eqnarray}
&&S(\sum_{\gamma =1}^n\frac 1{p_\gamma }|\gamma \rangle \langle \gamma |\rho
^{AB}|\gamma \rangle \langle \gamma |)-S(\sum_{\gamma =1}^n\frac 1{p_\gamma
}|\gamma \rangle \langle \gamma |\rho ^A|\gamma \rangle \langle \gamma |)
\nonumber \\
&&\geq \sum_{\gamma =1}^n\langle \gamma |\rho ^A|\gamma \rangle S(\frac{%
|\gamma \rangle \langle \gamma |\rho ^{AB}|\gamma \rangle \langle \gamma |}{%
p_\gamma \langle \gamma |\rho ^A|\gamma \rangle })-\sum_{\gamma =1}^n\langle
\gamma |\rho ^A|\gamma \rangle S(\frac{|\gamma \rangle \langle \gamma |}{%
p_\gamma })  \nonumber \\
&&=\sum_{\gamma =1}^n\langle \overline{\gamma }|\rho ^A|\overline{\gamma }
\rangle S(\frac{|\overline{\gamma }\rangle \langle \overline{\gamma }|\rho
^{AB}|\overline{\gamma }\rangle \langle \overline{\gamma }|}{\langle
\overline{\gamma }|\rho ^A|\overline{\gamma }\rangle }).  \nonumber
\end{eqnarray}
Where we have used $p_\gamma =\langle \gamma |\gamma \rangle $, $\langle \gamma |\rho ^A|\gamma \rangle =\langle \overline{\gamma }|\rho ^A|\overline{\gamma }\rangle =tr_B\langle \gamma |\rho ^{AB}|\gamma \rangle =tr_B\langle \overline{\gamma} |\rho ^{AB}|\overline{\gamma} \rangle $, $S(\frac{|\gamma \rangle \langle \gamma |}{p_\gamma })=0$, and $S(\frac{|\overline{\gamma }\rangle \langle \overline{\gamma }|\rho ^{AB}|\overline{\gamma }\rangle \langle \overline{\gamma }|}{\langle \overline{\gamma }|\rho ^A|\overline{\gamma }\rangle })=S(\frac{\langle \overline{\gamma }|\rho ^{AB}|\overline{\gamma }\rangle }{\langle \overline{\gamma }|\rho ^A|\overline{\gamma }\rangle })$. This leads to Eq.(26), and end the proof.

To find the states for $D_A^R(\rho ^{AB})=0$ or  $\overline{D}_A^R(\rho ^{AB})=0$, we make a digression to introduce an elegant result in \cite{Koashi2004}. Given a tripartite pure state $\rho ^{ABC}=|\psi \rangle \langle \psi |$ of a joint system ABC, by Schmidt decomposition (\cite{Nielson2000}, 2.5) we write $\rho ^{ABC}$ as
\begin{eqnarray}
\rho ^{ABC}=\sum_{i,j=1}^m\sqrt{p_ip_j}|\psi _i^A\rangle \langle \psi _j^A|\otimes |\psi _i^{BC}\rangle \langle \psi _j^{BC}|,
\end{eqnarray}
where $m=rank \rho ^{BC}\leq n_A$, $p_i>0$, $\sum_ip_i=1$, $\{|\psi _i^A\rangle \}$, $\{|\psi _i^{BC}\rangle \}$ are orthonormal sets of $H^A$ and $H^B\otimes H^C$, respectively. Performing a Neumark extension measurement $\{|\overline{\gamma }\rangle \langle \overline{\gamma }|\}_{\gamma =1}^n$ on system A, then
\begin{eqnarray}
\widetilde{\rho _\gamma ^B}=tr_C[(\sum_{i=1}^n\langle \overline{\gamma }|\psi _i^A\rangle \sqrt{p_i}|\psi _i^{BC}\rangle )(\sum_{j=1}^n\langle \psi _j^A|\overline{\gamma }\rangle \sqrt{p_j}\langle \psi _j^{BC}|)].
\end{eqnarray}	
Notice that $\{\sum_{i=1}^n\langle \overline{\gamma }|\psi _i^A\rangle \sqrt{p_i}|\psi _i^{BC}\rangle \}_\gamma $ is just a pure state decomposition of $\rho ^{BC}$ (\cite{Nielson2000}, 2.4.2), and all Neumark extension measurements realize all pure state decompositions of $\rho ^{BC}$, this leads to the result in \cite{Koashi2004}
\begin{eqnarray}
E(\rho ^{BC})=\inf_{R \otimes I_B}\sum_\alpha p_\alpha S(\widetilde{\rho _\alpha ^B}/p_\alpha ),
\end{eqnarray}
where $E(\rho ^{BC})$ is the entanglement of formation (EOF) \cite{Bennett1996} of $\rho ^{BC}$.

Here, we make an explanation. From Eq.(27), $\rho ^{BC}=\sum_{i=1}^mp_i|\psi _i^{BC}\rangle \langle \psi _i^{BC}|$ is the eigendecomposition of $\rho ^{BC}$, so all pure state decompositions of $\rho ^{BC}$ are $F=\cup _{N\geq m}F_N,$ where (\cite{Nielson2000}, 2.4.2)
\begin{eqnarray}
\fl
F_N=\{\{\sum_{i=1}^NU_{\lambda i}p_i|\psi _i^{BC}\rangle \}_{\lambda =1}^N:p_i=0 \ for \ i\geq m;U=(U_{\lambda i})\ is \ any \ N \times N \ unitary \ matrix \}. \nonumber
\end{eqnarray}
Note that if $m\leq N\leq N_1$, then $F_N\subset F_{N_1}$. This is true since any $N\times N$ unitary matrix multiplied by a $(N_1-N)\times (N_1-N)$ identity matrix becomes an $N_1\times N_1$ unitary matrix. Since $n\geq n_A\geq m$, that is why all Neumark extension measurements realize all pure state decompositions of $\rho ^{BC}$ in Eq.(28).

Now if we add $S(\rho ^A)-S(\rho ^{AB})$ to both sides of Eq.(29), we obtain
\begin{eqnarray}
\overline{D}_A^R(\rho ^{AB})=E(\rho ^{BC})+S(\rho ^A)-S(\rho ^{AB}).
\end{eqnarray}

Notice that $\overline{D}_A^R(\rho ^{AB})$ and $E(\rho ^{BC})$ can be achieved by the same Neumark extension measurement. It is known $E(\rho ^{BC})$  can be achieved by a finite $l$ $(l\geq m)$ pure decomposition \cite{Uhlmann1998}, then, correspondingly, $\overline{D}_A^R(\rho ^{AB})$ can be achieved by a Neumark extension measurement $\{|\overline{\gamma }\rangle \langle \overline{\gamma }|\}_{\gamma =1}^n$ with finite $n$. Moreover, if $E(\rho ^{BC})$ can be achieved by a finite $l$ ($m\leq l\leq n_A$) pure decomposition, then $\overline{D}_A^R(\rho ^{AB})$ can be achieved by a projective measurement, i.e., $\overline{D}_A^R(\rho ^{AB})=D_A^P(\rho ^{AB})$. From these facts we obtain proposition 8 and proposition 9 below.

\textit{Proposition 8.}
Quantum discord $D_A^P(\rho ^{AB})$ defined in Eq.(1) of any state $\rho ^{AB}$ of $n_A\times 2$ systems with rank no more than 2 can be analytically obtained according to Eq. (30).

\textit{Proof.} Suppose $\rho ^{AB}=\sum_{i=1}^2q_i|\psi _i^{AB}\rangle \langle \psi _i^{AB}|,$ $q_i\geq 0$, $\sum_{i=1}^2q_i=1$, $\{|\psi _i^{AB}\rangle \}_{i=1}^2$is an orthonormal set in $H^A\otimes H^B$. We purify  $\rho ^{AB}$ as $|\psi \rangle =\sum_{i=1}^2\sqrt{q_i}|\psi _i^{AB}\rangle |\psi _i^C\rangle $, where $\{|\psi _i^C\rangle \}_{i=1}^2$ is an orthonormal set in $H^C$. At the same time we can also write $|\psi \rangle $ as $|\psi \rangle =\sum_{i=1}^m\sqrt{p_i}|\psi _i^A\rangle |\psi _i^{BC}\rangle $ by schmidt decomposition if we regard BC as one system, where $m=rank\rho ^{BC}\leq \min \{n_A,4\}$, $p_i\geq 0$, $\sum_{i=1}^4p_i=1$, $\{|\psi _i^A\rangle \}_{i=1}^4$ and $\{|\psi _i^{BC}\rangle \}_{i=1}^4$ are orthonormal sets in $H^A$ and $H^B\otimes H^C$ respectively. Because $E(\rho ^{BC})$ allows analytical expression for any two qubits state $\rho ^{BC}$ \cite{Wootters1998}, and $E(\rho ^{BC})$ can be achieved by an m-vector pure decomposition of $\rho ^{BC}$, so $D_A^R(\rho ^{AB})$ can be achieved by an m-dimensional projective measurement (on the space spanned by $\{|\psi _i^A\rangle \}_{i=1}^m$) and it can be extended to a projective measurement since $m\leq n_A$. Hence $\overline{D}_A^R(\rho ^{AB})=D_A^P(\rho ^{AB})$ and we complete this proof.

\textit{Proposition 9.} The quantum discords $D_A^P(\rho ^{AB})$, $D_A^R(\rho ^{AB})$, $\overline{D}_A^R(\rho ^{AB})$ of a bipartite state $\rho ^{AB}$ defined in Eq.(1), Eq.(11) and Eq.(16), hold that
\begin{equation}
D_A^{P}(\rho ^{AB})=0\Longleftrightarrow D_A^{R}(\rho
^{AB})=0\Longleftrightarrow \overline{D}_A^R(\rho ^{AB}).
\end{equation}

\textit{Proof.} Suppose  $\overline{\overline{D}}_A^N(\rho ^{AB})$ can be achieved by an element $\{|\overline{\gamma }\rangle \langle \overline{\gamma }|\}_{\gamma =1}^n$ of finite $n$ in the set $N$. Therefore, similar to Eq. (2), we have
	\begin{eqnarray}
 \overline{\overline{D}}_A^N(\rho ^{AB})=0\Longleftrightarrow \rho ^{AB}=\sum_{\gamma =1}^np_\gamma |\overline{\gamma }\rangle \langle \overline{\gamma }|\otimes \rho _\gamma ^B,
	\end{eqnarray}
where $\{|\overline{\gamma }\rangle \}_{\gamma =1}^n$ is an arbitrary orthonormal set in $H_n^A$, $p_\gamma \geq 0$, $\sum_{\gamma =1}^np_\gamma =1$, $\rho _\gamma ^B$ are density operators on $H^B$. But $\{|\overline{\gamma }\rangle \}_{\gamma =1}^n$ is actually in $H^A$ since $\rho ^{AB}$ is on $H^A\otimes H^B$.  As a result,
\begin{eqnarray}
\overline{\overline{D}}_A^N(\rho ^{AB})=0\Longleftrightarrow D_A^P(\rho ^{AB})=0.
\end{eqnarray}	
Combining Eq.(26), we obtain Eq.(31).

\section{Summary}

We investigated some generalizations of quantum discord
which were defined over general measurements, rank-1 general measurements or Neumark extension measurements. The nonnegativity and zero-discord states were emphasized and some relations about different quantum discords were discussed.

In quantum information and quantum computation, we aim for an exquisite
level of control over the measurements, so it is natural to consider the
more comprehensive general measurements (such as the optimal way to distinguish a set of quantum states)
rather than projective measurements. We expect that these discussions about the generalizations of quantum discord may provide more extensive understandings for
characterizations of the nonlocal correlation.

\section*{Acknowledgements}

This work was supported by National Natural Science Foundation of China
(Grant Nos. 10775101). The author thanks Qing Hou, Xuewen Liu, Zhao Kang and
Yingfeng Xu for helpful discussions.

\end{document}